\documentclass[english,aps,pre,groupedaddress,showpacs,preprint]{revtex4}
\usepackage[T1]{fontenc}
\usepackage[latin1]{inputenc}
\usepackage{graphicx}
\usepackage{amssymb}

\makeatletter

\usepackage{amsbsy}
\newcommand{\binom}[2]{{#1 \choose #2}}
\makeatletter
\makeatletter
\usepackage{amsfonts}
\usepackage{graphics}
\usepackage{fixmath}
\usepackage{mathrsfs}
\usepackage{epsfig}
\newcommand{\prs}[1]{{\left(#1\right)}}

\newcommand{\bS}{\mathbf{S}}
\newcommand{\bw}{\mathbf{w}}

\newcommand{\intoo}{{\int_{-\infty}^{\infty}}}
\newcommand{\hmu}{{\hat{\mu}}}

\makeatother

\usepackage{babel}
\makeatother

\begin{document}

\title{Can a student learn optimally from two different teachers?}

\author{J. P. Neirotti}

\affiliation{Aston University, the Neural Computing Research Group, Birmingham,
United Kingdom}

\begin{abstract}
We explore the effects of over-specificity in learning algorithms
by investigating the behavior of a student, suited to learn optimally
from a teacher $\mathbf{B}$, learning from a teacher $\mathbf{B}'\neq\mathbf{B}$.
We only considered the supervised, on-line learning scenario with
teachers selected from a particular family. We found that, in the
general case, the application of the optimal algorithm to the wrong
teacher produces a residual generalization error, even if the right
teacher is harder. By imposing mild conditions to the learning algorithm
form we obtained an approximation for the residual generalization
error. Simulations carried in finite networks validate the estimate
found.
\end{abstract}


\maketitle

\section{Introduction}

Neural networks are connectivist models inspired on the dynamical
behavior of the brain \citet{rumelhart}. They are not only theoretically
interesting models, they can also be used in a number of applications,
from voice recognition systems to curve fitting software. Probably
the properties that make neural networks most useful are their potentiality
to store patterns and their capability for learning tasks. 

One of the most well-studied types of networks is feed-forward. What
characterizes a feed-forward network is that the flux of information
follows a non-loopy path from input to output nodes, making the information
processing much faster. Perceptrons \citet{papert} are feed-forward
networks with no internal nodes and only one output; they have been
utilised for a number theoretical studies and applications of statistical
mechanics techniques \citet{engel}. In particular, the knowledge
of Hebbian learning algorithms in an on-line scenario is quite complete.

In the present article we study the ability of a student \textbf{J},
using an algorithm for learning optimally from a specific teacher
$\mathbf{B}$, to learn from a teacher $\mathbf{B}'$. If a student
is adapted to learn from a \emph{difficult }teacher, it is not unreasonable
to expect that it will be able to learn from an \emph{easier }one.
To formally analyze this problem we need to quantify the hardness
of the teachers, set up the scenario where the learning process would
take place and thus quantify the student's performance.

Attempts to quantify hardness as an inherent property of the observed
object have given origin to many formal definitions of complexity
\citet{church,turing0,hartmanis,kolmogorov,ming}. Recently \citet{fra-0}
L. Franco has proposed to quantify a (Boolean) function's hardness
by the size of the minimal set of examples needed to train a feed-forward
network, with a predetermined architecture until reaching zero prediction
error. He also found \citet{fra-1,fra-2} that in this minimal set
there are many pairs of examples that, although only differing in
a finite number \emph{$P=1,2,\dots$} of entries, they have different
outputs, implying that these examples are located at each side of
the classification boundary (similar to the support vectors for SVMs
\citet{ton}). Further investigation showed that the average discrepancy
of the function's outputs (measure over neighboring pairs) is correlated
to the generalization ability of the network implementing the function.
In order to contour the use of the neural network and its minimal
training set, Franco proposed to use the average distance sensitivity
directly as a measure of the function's hardness. This is probably
the most suitable measure for our study given that the nature of the
measure itself is linked to the concept of generalization ability. 

The hardness measure we will use is the average output discrepancy
taken over all pairs of inputs at a given Hamming distance \emph{P.
}Formally, for a given Boolean function $f:\left\{ \pm1\right\} ^{N}\to\left\{ \pm1\right\} ,$
the $P$th distance sensitivity component $\mathfrak{d}_{P}^{N}[f]$
is the functional \begin{equation}
\mathfrak{d}_{P}^{N}\left[f\right]=2^{-N}\sum_{\mathbf{S}\in\left\{ \pm1\right\} ^{N}}\binom{N}{P}^{-1}\sum_{\mathbf{S}^{\prime}\in\upOmega_{P}\left(\mathbf{S}\right)}\frac{1-f\left(\mathbf{S}\right)f\left(\mathbf{S}^{\prime}\right)}{2}\,,\label{eq:defpboolean}\end{equation}
 where $\upOmega_{P}\left(\mathbf{S}\right)=\left\{ \mathbf{S}^{\prime}\in\left\{ \pm1\right\} ^{N}|\sum_{j=1}^{N}\Theta\left(-S_{j}S_{j}^{\prime}\right)=P\right\} $.
$\upOmega_{P}(\mathbf{S})$ is the set of inputs $\mathbf{S}'$ that
differ from \textbf{S} in \emph{P} entries.

Dilution gives rise to networks with fewer connections, which can
be more efficient in solving tasks and can be more easily implemented
in hardware. Diluted perceptrons have been widely studied using statistical
mechanics techniques \citet{canning,bouten,jort,khulman,lopez,malzahn}
and have also been studied as an approximation to more difficult Boolean
functions \citet{kalai,odonnell}. Probably the most important features
of diluted perceptrons related to the present work are the existence
of analytical expressions for the sensitivity component (\ref{eq:defpboolean})
and the associated optimal learning algorithm (see below). 

Consider a perceptron characterized by a synaptic vector $\mathbf{B}^{(m)}\in\mathbb{R}^{N}$
that classifies binary vectors $\mathbf{S}\in\left\{ \pm1\right\} ^{N}$
with labels $\sigma_{\mathbf{B}}\in\left\{ \pm1\right\} $ according
to the rule $\sigma_{\mathbf{B}}=\mathrm{sgn}\left(\mathbf{B}^{(m)}\cdot\mathbf{S}\right).$
If $[\mathbf{B}^{(m)}]_{i}=\delta(i\in\mathbb{I}_{m})O(\sqrt{N/m})+\delta(i\notin\mathbb{I}_{m})o(\sqrt{m/N})$
where $\mathbb{I}_{m}\in\left\{ 1,2,\dots,N\right\} $ is a set of
\emph{m} (odd) different indexes $1\leq i\leq N$ we have a \emph{diluted
binary perceptron}. In our calculations we will consider $[\mathbf{B}^{(m)}]_{i}=\delta(i\in\mathbb{I}_{m})\,\sqrt{N/m}$\emph{
}where $m\ll N$ will be kept finite. 

For the binary perceptron $\mathbf{B}^{(m)}$ the distance sensitivity
component (\ref{eq:defpboolean}) in the large system limit ($P<N\to\infty$
with $p\equiv P/N<\infty$) $\mathfrak{d}^{(m)}(p)$ is given by (\ref{eq:dd})\begin{eqnarray*}
\mathfrak{d}^{(m)}(p) & = & \frac{1}{2}-\frac{1}{2}\sum_{n=0}^{(m-1)/2}a_{n}^{m}\,(1-2p)^{2n+1}\\
a_{n}^{m} & = & \frac{1}{4^{m-1}}{\displaystyle \,\binom{m}{2n+1}\,\left[\binom{2n}{n}\,\binom{m-1-2n}{(m-1)/2-n}\,\binom{(m-1)/2}{n}^{-1}\right]^{2}}.\end{eqnarray*}
As it is shown in the Appendix \ref{sec:Distance-sensitivity}, and
following \citet{odonnell}, $\mathfrak{d}^{(m)}(p)$ are a family
of concave functions, ordered according to $\mathfrak{d}^{(m)}(p)<\mathfrak{d}^{(m+2)}(p)\;\forall p\in(0,\frac{1}{2}).$
Therefore, the order given by the hardness measure coincides with
the order given by \emph{m}, thus the larger \emph{m} is the harder
the Teacher.

Another reason that appeals for using a diluted perceptrons as a teacher
is that it is possible to obtain the correspondent optimal learning
algorithm analytically. In a supervised on-line learning scenario,
the synaptic vector of the student perceptron \textbf{J} is adjusted
after receiving new information in the form of the pair $\left(\mathbf{S},\sigma_{\mathbf{B}}\right),$
following the rule\begin{equation}
\mathbf{J}_{\mathrm{new}}=\mathbf{J}_{\mathrm{old}}+F\,\frac{\sigma_{\mathbf{B}\mathrm{new}}\mathbf{S}_{\mathrm{new}}}{\sqrt{N}},\label{eq:heb}\end{equation}
where $\mathbf{J}\in\mathbb{R}^{N}$, $\sigma_{\mathbf{B}}=\mathrm{sgn}(\mathbf{B}\cdot\mathbf{S})$
is the classification given to the example by the teacher \textbf{B}
and \emph{F} is the learning amplitude or algorithm. The parameters
of the problem are\[
h\equiv\frac{\mathbf{J}\cdot\mathbf{S}}{|\mathbf{J}|},\qquad b\equiv\frac{\mathbf{B}\cdot\mathbf{S}}{|\mathbf{B}|},\qquad Q\equiv\frac{\mathbf{J}\cdot\mathbf{J}}{N},\qquad R\equiv\frac{\mathbf{B}\cdot\mathbf{J}}{|\mathbf{B}||\mathbf{J}|}\]
 where \emph{h} is known as the student's post-synaptic field, \emph{b}
is the teacher's post-synaptic field, i.e. $\mathrm{sgn}(b)=\sigma_{\mathbf{B}}$,
\emph{Q} is the normalized length of \textbf{J} and \emph{R} is the
overlap between teacher and student. 

Following \citet{engel} we found that the equation of motion for
the overlap \emph{R} in terms of the total number of examples received
$\alpha N$, in the large size limit $N\to\infty$, is\begin{equation}
\frac{\mathrm{d}R}{\mathrm{d}\alpha}=\left\langle \frac{F}{\sqrt{Q}}\left[\left\langle |b|\right\rangle _{b|\phi}-R\phi\right]-\frac{RF^{2}}{2Q}\right\rangle _{\phi},\label{eq:evolr}\end{equation}
 where $\left\langle \cdot\right\rangle _{\phi}$ represents an average
over the distribution $\mathcal{P}(\phi)$ and $\phi\equiv\sigma_{\mathbf{B}}h$.
The solution of this equation represents the evolution of the overlap
\emph{R} as a function of the \emph{time} $\alpha$. 

Remembering that the generalization error is defined as $e_{g}=\mathrm{arccos}(R)/\pi$,
and that the learning curve is the error as a function of $\alpha$,
we define the residual error as the asymptotic value of the learning
curve at large values of $\alpha$, i.e. $e_{g}^{\star}=\lim_{\alpha\to\infty}e_{g}(\alpha)$.

By the application of a variational technique it is possible to obtain
an expression for the optimal algorithm $F_{\mathrm{op}}$. The optimal
algorithm is the algorithm that produces the fastest decaying learning
curve and can be generically expressed as\[
F_{\mathrm{op}}=\frac{\sqrt{Q}}{R}\left[\left\langle |b|\right\rangle _{b|\phi}-R\phi\right].\]

\section{Analytical results}

Following \citet{caki} we can prove (see Appendix \ref{sec:Optimal-learning})
that, for a perceptron with dilution \emph{m} \begin{eqnarray}
\mathcal{P}(\phi|m) & = & \frac{1}{2^{m-1}}\sum_{k=0}^{(m-1)/2}\binom{m}{(m-1)/2-k}\,\mathcal{N}(\phi|R\mu_{k},1-R^{2})\label{eq:pp}\\
\left\langle |b|\right\rangle _{b|\phi,m} & = & \frac{\sum_{k=0}^{(m-1)/2}\binom{m}{(m-1)/2-k}\,\mu_{k}\,\mathcal{N}(\phi|R\mu_{k},1-R^{2})}{\sum_{k=0}^{(m-1)/2}\binom{m}{(m-1)/2-k}\,\mathcal{N}(\phi|R\mu_{k},1-R^{2})}\label{eq:bb}\\
F_{\mathrm{op}}^{(m)} & = & \frac{\sqrt{Q}}{R}\left[\frac{\sum_{k=0}^{(m-1)/2}\binom{m}{(m-1)/2-k}\,\mu_{k}\,\mathcal{N}(\phi|R\mu_{k},1-R^{2})}{\sum_{k=0}^{(m-1)/2}\binom{m}{(m-1)/2-k}\,\mathcal{N}(\phi|R\mu_{k},1-R^{2})}-R\phi\right],\label{eq:optimo}\end{eqnarray}
where $\mu_{k}=(2k+1)/\sqrt{m}$ and $\mathcal{N}(x|\mu,\sigma^{2})$
is a Normal distribution centered at $\mu$ with variance $\sigma^{2}$.
Observe that (\ref{eq:bb}) is needed for computing the evolution
(\ref{eq:evolr}), and (\ref{eq:optimo}) represents the optimal learning
algorithm. 

Suppose that the teacher is characterized by a dilution \emph{$m_{\mathbf{B}}$}
and the student implements an algorithm (\ref{eq:optimo}) for learning
a Teacher perceptron with dilution $m$. This is equivalent to having
prepared a student to learn optimally from $\mathbf{B}^{(m)}$ and
now exposing it to $\mathbf{B}^{(m_{\mathbf{B}})}\neq\mathbf{B}^{(m)}$.
Let us define the quantity\begin{equation}
\Upsilon(\phi|R,m)\equiv\left\langle \left|b\right|\right\rangle _{b|\phi,m}-R\phi.\label{eq:upsi}\end{equation}
In this settings, the algorithm has the form $F^{(m)}=\frac{\sqrt{Q}}{R}\Upsilon(\phi|R,m)$
and the distribution of $\phi$ is a function of $m_{\mathbf{B}}.$
The evolution of the overlap \emph{R} is given now by the equation
(\ref{eq:evolr})\[
\frac{\mathrm{d}R}{\mathrm{d}\alpha}=\left\langle \frac{1}{R}\Upsilon(\phi|R,m)\,\Upsilon(\phi|R,m_{\mathbf{B}})-\frac{1}{2R}\Upsilon^{2}(\phi|R,m)\right\rangle _{\phi|m_{\mathbf{B}}}\]
which can be reduced to\begin{eqnarray}
\frac{\mathrm{d}R^{2}}{\mathrm{d}\alpha} & = & 2\left\langle \Upsilon(\phi|R,m)\,\Upsilon(\phi|R,m_{\mathbf{B}})\right\rangle _{\phi|m_{\mathbf{B}}}-\left\langle \Upsilon^{2}(\phi|R,m)\right\rangle _{\phi|m_{\mathbf{B}}}\label{eq:problem00}\\
 & = & \left\langle \Upsilon^{2}(\phi|R,m_{\mathbf{B}})\right\rangle _{\phi|m_{\mathbf{B}}}-\left\langle \left[\Upsilon(\phi|R,m_{\mathbf{B}})-\Upsilon(\phi|R,m)\right]^{2}\right\rangle _{\phi|m_{\mathbf{B}}}.\label{eq:problem}\end{eqnarray}
The overlap \emph{R} grows from zero to a stationary value, thus we
expect the second term at the RHS of (\ref{eq:problem}) to be smaller
than the first one. In the asymptotic regime ($\alpha\to\infty$)
the derivative is zero, implying that no further changes are expected
in the overlap, and then we have that\begin{equation}
\left\langle \Upsilon^{2}(\phi|R^{\star},m_{\mathbf{B}})\right\rangle _{\phi|m_{\mathbf{B}}}=\left\langle \left[\Upsilon(\phi|R^{\star},m)-\Upsilon(\phi|R^{\star},m_{\mathbf{B}})\right]^{2}\right\rangle _{\phi|m_{\mathbf{B}}}\label{eq:prob1}\end{equation}
 where $R^{\star}\equiv\lim_{\alpha\uparrow\infty}R(\alpha)$.

\begin{figure}
\includegraphics[scale=0.7]{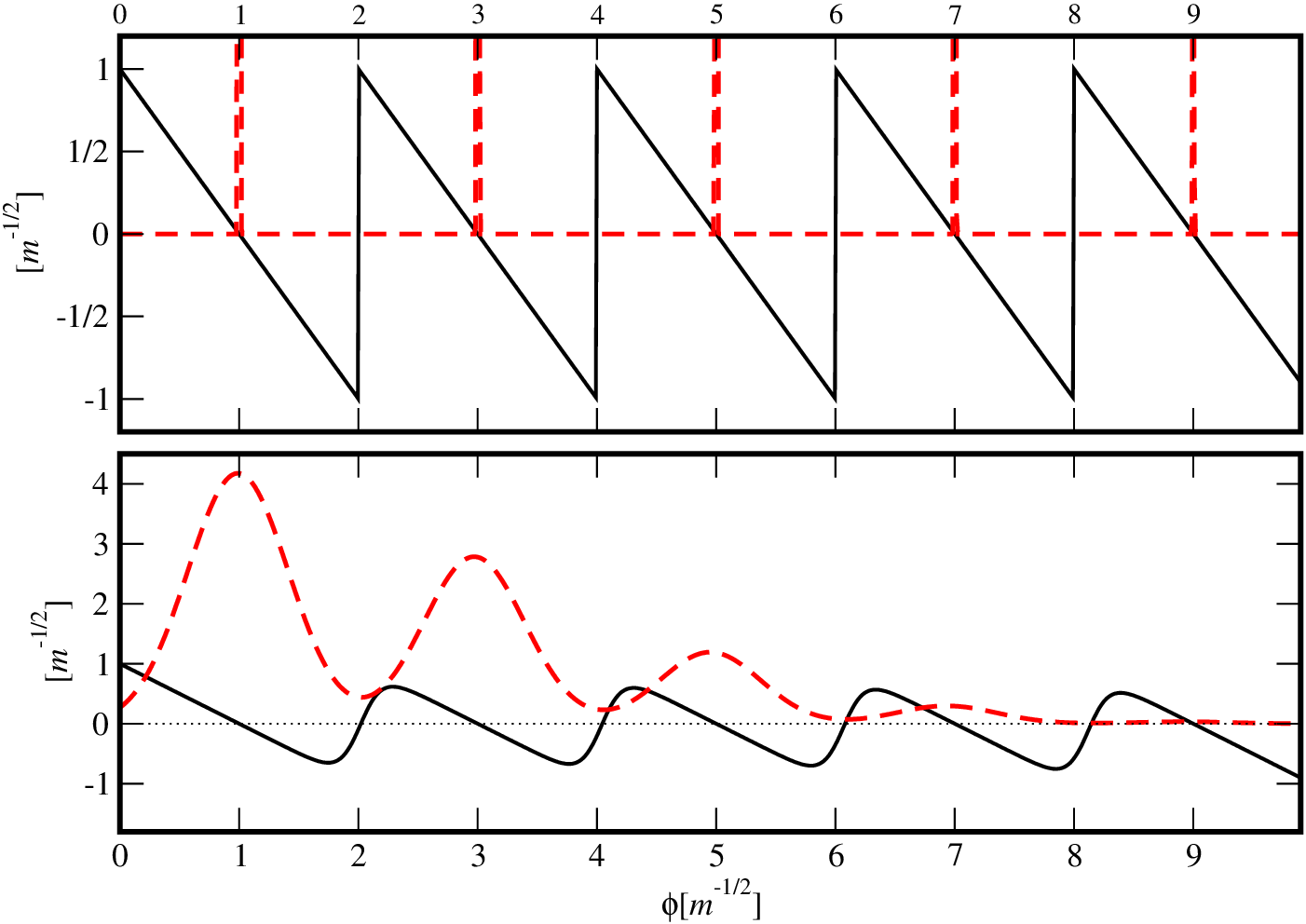}

\caption{$\Upsilon(\phi|R,m)$ (full curve) and the probability of the of the
stability $\mathcal{P}(\phi|m)$ (dashed curve) against $\phi$ in
units of $1/\sqrt{m}$ for \emph{$R=1$ }(upper panel) and $R=0.99$
(lower panel) for $m=9.$ Observe that for $R=1$ (upper panel) the
average of the LHS (\ref{eq:prob1}) involves only the points at which
$\Upsilon(\phi|1,m)$ is zero, whilst for $R<1$ (lower panel) the
same average requires a more intensive calculation.\label{fig:1}}

\end{figure}

Observe that if $m=m_{\mathbf{B}},$ the second term of the RHS of
(\ref{eq:problem00}) is zero, the algorithm applied is optimal and
the overlap reaches $R^{\star}=1$ with the smallest possible set
of examples. If perfect learning implies $R^{\star}=1$ it is natural
to ask for what values of \emph{m} the student can learn a teacher
with dilution $m_{\mathbf{B}}$ without errors. From (\ref{eq:pp})
and (\ref{eq:bb}) we have that, for $R=1,$\begin{eqnarray}
\mathcal{P}(\phi|m_{\mathbf{B}}) & = & \frac{\sqrt{m_{\mathbf{B}}}}{2^{m_{\mathbf{B}}-1}}\sum_{k=0}^{(m_{\mathbf{B}}-1)/2}\binom{m_{\mathbf{B}}}{(m_{\mathbf{B}}-1)/2-k}\,\delta\left(\sqrt{m_{\mathbf{B}}}\,\phi-(2k+1)\right)\label{eq:1}\\
\Upsilon(\phi|1,m_{\mathbf{B}}) & = & \frac{1}{\sqrt{m_{\mathbf{B}}}}\left[1+\sum_{k=1}^{(m_{\mathbf{B}}-1)/2}\Theta\left(\sqrt{m_{\mathbf{B}}}\,\phi-2k\right)\right]-\phi\label{eq:2}\\
\Upsilon(\phi|1,m) & = & \frac{1}{\sqrt{m}}\left[1+\sum_{k=1}^{(m-1)/2}\Theta\left(\sqrt{m}\,\phi-2k\right)\right]-\phi.\label{eq:3}\end{eqnarray}
 The LHS of (\ref{eq:prob1}), averaged over (\ref{eq:1}) is zero
(see figure \ref{fig:1}). This is due to the fact that $\Upsilon((2k+1)/\sqrt{m_{\mathbf{B}}}|1,m_{\mathbf{B}})=0$.
Therefore, in order to satisfy (\ref{eq:prob1}) we also need that
$\Upsilon((2k+1)/\sqrt{m_{\mathbf{B}}}|1,m)=0$. Particularly, for
$k=0$ these two equation imply that\begin{eqnarray*}
\sqrt{\frac{m}{m_{\mathbf{B}}}} & = & 1+\sum_{k=1}^{(m-1)/2}\Theta\left(\sqrt{\frac{m}{m_{\mathbf{B}}}}-2k\right).\end{eqnarray*}
Therefore \begin{equation}
\sqrt{\frac{m}{m_{\mathbf{B}}}}=2q+1,\label{eq:solR1}\end{equation}
 where \emph{q} is a suitable, non-negative integer. Thus, the condition
for $R=1$ to be a solution of (\ref{eq:prob1}) is that there exist
$q\in\mathbb{N}\cup\{0\}$ such that $m=(2q+1)^{2}m_{\mathbf{B}}.$ 

If this is not true, the solution of (\ref{eq:prob1}) is at $R^{\star}<1.$
We will present an approach based on the assumption that the root
$R^{\star}$ occurs in a regime where the Gaussian distributions $\mathcal{N}(\phi|R^{\star}\mu_{k},1-R^{\star2})$
in (\ref{eq:pp}) and (\ref{eq:bb}) have a small overlap. This could
be ensured if the separation of two adjacent Gaussian components were
larger than two standard deviations, i.e.\begin{eqnarray}
R^{\star}|\mu_{k}-\mu_{k+1}|=\frac{2R^{\star}}{\sqrt{m}} & \gg & 2\sqrt{1-R^{\star2}}\label{eq:approx1}\\
1 & \gg & m\,\frac{1-R^{2}}{R^{2}}\label{eq:approx}\end{eqnarray}
 At $R=1$ the curve $\Upsilon(\phi|1,m)$ is discontinuous at $\phi\equiv2k/\sqrt{m}$
and the probability $\mathcal{P}(\phi|m)$ is a linear combination
of delta functions centered at $\phi=(2k+1)/\sqrt{m}$ (upper panel
of figure \ref{fig:1}). For $R<1$ (figure \ref{fig:1}, lower panel),
$\Upsilon(\phi|R,m)$ is continuous and $\mathcal{P}(\phi|m)$ is
a linear combination of Gaussian distributions centered at $\phi=R(2k+1)/\sqrt{m}$
with variance $1-R^{2}$. In both cases $\Upsilon(\phi|R,m)$ appears
to be a periodic function of $\phi$ with period $\phi_{T}\equiv2R/\sqrt{m},$
in the support of $\mathcal{P}(\phi|m)$ $\mathscr{D}_{\phi}\subset\mathbb{R},$
i.e.\begin{eqnarray*}
\Upsilon(\phi|R,m) & \simeq & \Upsilon\left(\phi+n\phi_{T}|R,m\right),\end{eqnarray*}
 and particularly for $R=1$ we have that\begin{equation}
\Upsilon(\phi|1,m)=\sum_{\ell=0}^{(m-1)/2}\Theta\left[\left(2(\ell+1)-\sqrt{m}\,\phi\right)\left(\sqrt{m}\,\phi-2\ell\right)\right]\left(\frac{2\ell+1}{\sqrt{m}}-\phi\right).\label{eq:sharp}\end{equation}
We can approximate $\Upsilon(\phi|R,m)$ by a suitable superposition
of Normal distributions. Consider the superposition\begin{equation}
\tilde{\Upsilon}(\phi|R,m)\equiv\int_{-\infty}^{\infty}\mathrm{d}r\, g(r)\,\mathcal{N}(\phi|r,1-R^{2}).\label{eq:model}\end{equation}
To determine the function $g(r)$, we perform a variational calculation
to minimize the error functional \[
\varepsilon[g]\equiv\frac{1}{2}\int_{\mathscr{D}_{\phi}}\mathrm{d}\phi\,\left[\Upsilon(\phi|R,m)-\tilde{\Upsilon}(\phi|R,m)\right]^{2}.\]
 Observe that the optimal function $g_{o}(r)$ is the solution of
the equation ${\displaystyle \left.\frac{\delta\varepsilon}{\delta g}\right|_{g_{o}}=0,}$
which implies that for all $r_{0}\in\mathbb{R}$ we have that \[
\int_{\mathscr{D}_{\phi}}\mathrm{d}\phi\,\left[\Upsilon(\phi|R,m)-\tilde{\Upsilon}(\phi|R,m)\right]\,\mathcal{N}(\phi|r_{0},1-R^{2})=0,\]
 in particular if $R=1$ (we assume that $g_{o}(r)$ is independent
of $R$)\begin{eqnarray*}
0 & = & \int_{\mathscr{D}_{\phi}}\mathrm{d}\phi\,\left[\Upsilon(\phi|1,m)-\int_{-\infty}^{\infty}\mathrm{d}r\, g_{o}(r)\,\delta(\phi-r)\right]\,\delta(\phi-r_{0})\\
g_{o}(r_{0}) & = & \Upsilon(r_{0}|1,m)\end{eqnarray*}
Therefore\begin{eqnarray}
\Upsilon(\phi|R,m) & \simeq & \int_{\mathscr{D}_{\phi}}\mathrm{d}r\,\Upsilon(r|1,m)\,\mathcal{N}(\phi|r,1-R^{2})\nonumber \\
 & = & \sum_{\ell=0}^{(m-1)/2}\int_{2\ell R/\sqrt{m}}^{2(\ell+1)R/\sqrt{m}}\mathrm{d}r\,\left[(2\ell+1)\frac{R}{\sqrt{m}}-r\right]\,\mathcal{N}(\phi|r,1-R^{2})\nonumber \\
 & = & \frac{R^{2}}{m}\int_{-1}^{1}\mathrm{d}t\, t\,\sum_{\ell=0}^{(m-1)/2}\mathcal{N}(\phi|R(2\ell+1-t)/\sqrt{m},1-R^{2}).\label{eq:ffi}\end{eqnarray}

Let us define the integral\begin{eqnarray}
\mathscr{I}_{m_{1},m_{2}} & \equiv & \int\mathrm{d}\phi\,\mathcal{P}(\phi|m_{\mathbf{B}})\,\Upsilon(\phi|R,m_{1})\,\Upsilon(\phi|R,m_{2}).\label{eq:int}\end{eqnarray}
 Following the development of Appendix \ref{sec:Derivation} we have
that \begin{equation}
\mathscr{I}_{m_{1},m_{2}}\simeq1-R^{2}\left[1-\frac{1}{2^{m_{\mathbf{B}}-1}}\sum_{k=0}^{(m_{\mathbf{B}}-1)/2}\binom{m_{\mathbf{B}}}{(m_{\mathbf{B}}-1)/2-k}\,\delta_{m_{1},k}^{\star}\delta_{m_{2},k}^{\star}\right],\label{eq:im1m2mb}\end{equation}
 where $\delta_{m_{j},k}^{\star}$ are given by \begin{equation}
\delta_{m_{j},k}^{\star}=\frac{2}{\sqrt{m_{j}}}\,\left[\!\!\left|\sqrt{\frac{m_{j}}{m_{\mathbf{B}}}}\,\frac{2k+1}{2}-\frac{1}{2}\right|\!\!\right]+\frac{1}{\sqrt{m_{j}}}-\frac{2k+1}{\sqrt{m_{\mathbf{B}}}},\label{eq:delcritico}\end{equation}
where $[\!|r|\!]$ is the closest integer to $r\in\mathbb{R}$.

Observe that from (\ref{eq:problem00}) we have that in the asymptotic
regime\begin{eqnarray*}
0 & = & 2\left\langle \Upsilon(\phi|R^{\star},m_{\mathbf{B}})\,\Upsilon(\phi|R^{\star},m)\right\rangle _{\phi|m_{\mathbf{B}}}-\left\langle \Upsilon^{2}(\phi|R^{\star},m)\right\rangle _{\phi|m_{\mathbf{B}}}\\
 & = & 2\mathscr{I}_{m_{\mathbf{B}},m}-\mathscr{I}_{m,m}\,\,,\end{eqnarray*}
 and observing that $\mathscr{I}_{m_{\mathbf{B}},m}=1-R^{2}$ (given
that $\delta_{m_{\mathbf{B}},k}^{\star}=0\;\forall k$) and \[
\mathscr{I}_{m,m}=1-R^{2}+\frac{R^{2}}{2^{m_{\mathbf{B}}-1}}\sum_{k=0}^{(m_{\mathbf{B}}-1)/2}\binom{m_{\mathbf{B}}}{(m_{\mathbf{B}}-1)/2-k}\,\delta_{m,k}^{\star2}\]
 then\begin{eqnarray}
R^{\star2} & = & \left[1+\frac{1}{2^{m_{\mathbf{B}}-1}}\sum_{k=0}^{(m_{\mathbf{B}}-1)/2}\binom{m_{\mathbf{B}}}{(m_{\mathbf{B}}-1)/2-k}\,\delta_{m,k}^{\star2}\right]^{-1}\label{eq:result}\end{eqnarray}
 and observe that $\delta_{m,k}^{\star}=0$ iff $m=(2q+1)^{2}m_{\mathbf{B}},\; q\in\mathbb{N},$
which is consistent with (\ref{eq:solR1}).

\section{Numerical Results}

\begin{figure}
\includegraphics[scale=0.7]{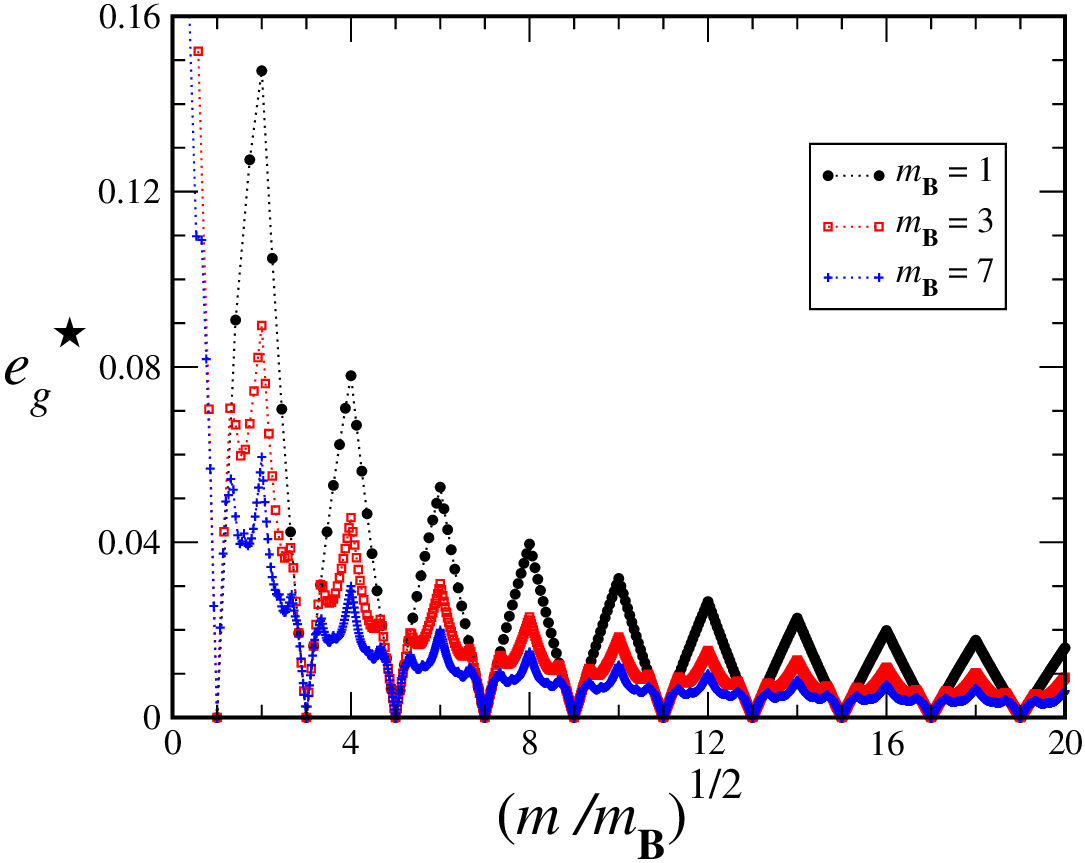}

\caption{Generalization error in the asymptotic regime $e_{g}^{\star}$ as
a function of $\sqrt{m/m_{\mathbf{B}}},$ for $m_{\mathbf{B}}=1,3,7$.
We have use (\ref{eq:result}) to compute the overlap \emph{$R^{\star}$.\label{fig:2}}}

\end{figure}

Using (\ref{eq:result}) we plot $e_{g}^{\star}=e_{g}(R^{\star})$
as a function of $\sqrt{m/m_{\mathbf{B}}}$ (see figure \ref{fig:2}).

\begin{figure}
\includegraphics[scale=0.7]{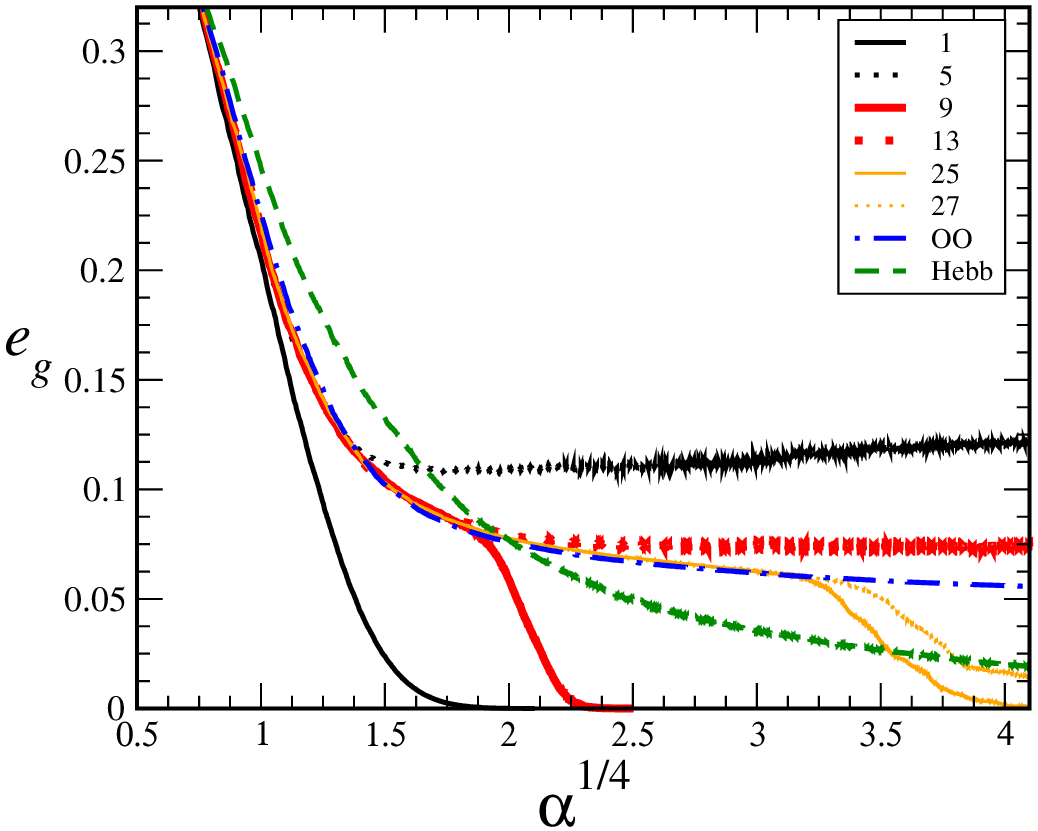}

\caption{Generalization error as a function of $\alpha^{\frac{1}{4}}$, for
a teacher with dilution $m_{\mathbf{B}}=1$ and students with $m=1,5,9,13,25,27$,
for a network with $N=51$. The curves that corresponds to the Hebb
algorithm ($F=1$, long dashed) and $m=\infty$ (dot dashed) are presented
as a reference. \label{fig:Galfa}}

\end{figure}

To validate our result shown in (\ref{eq:result}) we run a series
of numerical experiments consisting of a student learning from a Teacher
with only one bit ($m_{\mathbf{B}}=1$). The student updates its synaptic
vector following (\ref{eq:heb}) using a learning algorithm given
by (\ref{eq:optimo}) with $m=1,3,\dots,N.$ To compute the generalization
error we average over 50 realizations of the learning curve. The maximum
number of examples considered was $16\,000.$ In figure \ref{fig:Galfa}
we present the $e_{g}$ as a function of $\alpha^{\frac{1}{4}}$ for
$m=1,5,9,13,25,27$ and network size $N=51.$ We have chosen the exponent
$\frac{1}{4}$ to better show the curve features at short times and
the approach to the asymptotic regime. It is clear from the picture
that for $m=1^{2},3^{2},5^{2}$ the generalization error for large
$\alpha$ drops to zero as predicted. In order to extract the asymptotic
behaviour of the curves we applied the Bulirsch-Stoer algorithm \citet{bs}. 

\begin{figure}
\includegraphics[scale=0.7]{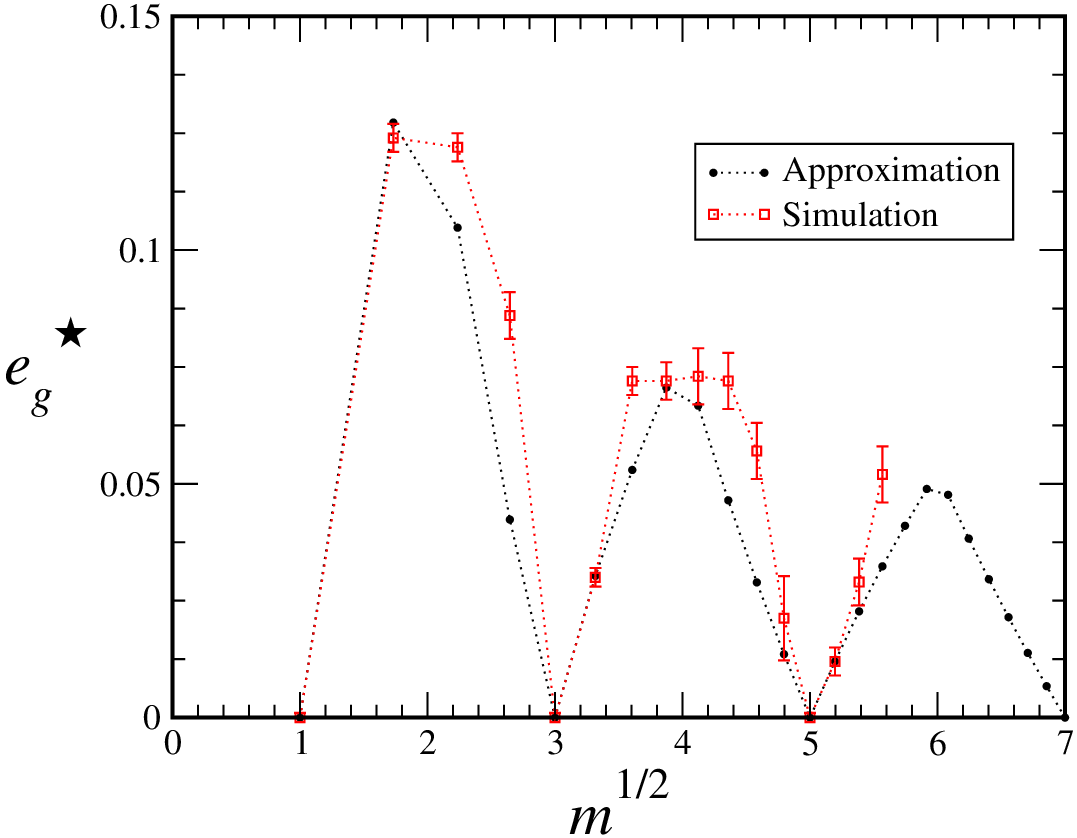}

\caption{Comparison of the asymptotic value of the generalization error using
(\ref{eq:result}) and the extrapolated values of the curves\emph{
}presented in figure \ref{fig:Galfa}\emph{.\label{fig:comp}}}

\end{figure}

In figure \ref{fig:comp} we present the extrapolated values of the
learning curves together with the values estimated by the application
of (\ref{eq:result}) as a function of $\sqrt{m}$. The error bars
are estimates obtained also by the application of the Bulirsch-Stoer
algorithm.

\section{Conclusions}

We studied the generalization capabilities of a student optimally
adapted for learning from a teacher \textbf{B}, when learning from
a teacher $\mathbf{B}'\neq\mathbf{B}.$ We observed that, although
the algorithm the student uses may be suited for learning from a harder
teacher, (as defined by Franco) that does not guarantee the success
of the process, as revealed by (\ref{eq:result}). This behavior is
due to the extreme specialization implied by the algorithm (\ref{eq:optimo}).
When this algorithm (with parameter \emph{m}) is applied to learn
from a teacher with $m_{\mathbf{B}}<m$, the student tries to extract
information from bits that the teacher does not use for producing
the correct classification. These interference effects produce mostly
bad results, originating a residual error in the asymptotic regime.
In this sense, the algorithm $F_{\mathrm{op}}^{(m)}$ is worse than
the Hebb algorithm $F_{\mathrm{Hebb}}=1.$

Despite the discrepancies shown in figure \ref{fig:comp}, our estimate
(\ref{eq:result}) reproduces faithfully the qualitative behaviour
observed in the simulations. There are two sources of uncertainty
that may account for the observed discrepancies: the (finite) size
of the network used and a not sufficiently large $\alpha.$ 

From figure \ref{fig:2}, the algorithm obtained by taking the limit
$m\to\infty$ in (\ref{eq:optimo})\[
F_{\mathrm{op}}^{(\infty)}=\sqrt{\frac{Q\,\left(1-R^{2}\right)}{2\pi\, R^{2}}}\;\frac{\exp\left(-{\displaystyle \frac{1}{2}\,\frac{R^{2}\phi^{2}}{1-R^{2}}}\right)}{\mathcal{H}\left(-{\displaystyle \frac{R\phi}{\sqrt{1-R^{2}}}}\right)},\]
where $\mathcal{H}(x)=\int_{x}^{\infty}\mathrm{d}y\,\mathrm{e}^{-\frac{y^{2}}{2}}\,/\sqrt{2\pi}$;
as reported by \citet{caki}, produces zero residual error for all
$m_{\mathbf{B}}$. The Hebb algorithm $F_{\mathrm{Hebb}}=1$ also
produces learning curves with zero residual error. In figure \ref{fig:Galfa}
we observe that the Hebb algorithm performs better than $F_{\mathrm{op}}^{(\infty)}$.
This is not a contradictory result. $F_{\mathrm{op}}^{(\infty)}$
is the algorithm that has the best average performance considering
a homogeneous distribution of teachers over the \emph{N}-sphere. For
a measure zero subset of vectors embedded in the \emph{N}-sphere,
like the perceptrons with finite dilution \emph{m}, $F_{\mathrm{op}}^{(\infty)}$
could perform worse than the Hebb algorithm, as it seems to be the
case here.

In order to obtain the fastest decaying learning curve, a student
has to infer the correct dilution of the teacher for choosing the
appropriate learning algorithm. Developing an efficient technique
for inferring the correct dilution parameter will be the subject of
our future research.

\begin{acknowledgments}
I would like to acknowledge the fruitful discussions with Dr L. Rebollo-Neira,
Dr R. C. Alamino, Dr L. Franco, Dr C. M. Ju\'arez and Prof. N. Caticha
which have enriched the contents of this article. 
\end{acknowledgments}

\appendix

\section{Distance sensitivity\label{sec:Distance-sensitivity}}

\textbf{S} and $\mathbf{S}'$ are vectors that differ in exactly \emph{P}
bits, i.e. $\sum_{j=1}^{P}\Theta(-S_{\sigma_{j}}S_{\sigma_{j}}^{\prime})=P$.
Taken \textbf{S} as a reference, we can construct a \emph{P}th neighbor
$\mathbf{S}'$ by choosing without replacement \emph{P} indexes from
1 to \emph{N} and flipping the correspondent entries in \textbf{S}.
There are $\binom{N}{P}$ different ways to choose \emph{P} indexes,
each one creating a different set of indexes $\mathbb{I}_{P}$. Introducing
the scaled variables $\mu=\bw\cdot\bS/\sqrt{N}$ and $\mu'=\bw\cdot\bS'/\sqrt{N}$
by means of Dirac delta functions and adding up over all possible
configurations \textbf{S}, we can express the discrepancy component
as\begin{eqnarray}
\mathfrak{d}_{P}^{N}(\mathbf{B}^{(m)}) & = & \binom{N}{P}^{-1}\intoo\frac{\mathrm{d}\mu\,\mathrm{d}\hmu}{2\pi}\,\frac{\mathrm{d}\mu'\,\mathrm{d}\hmu'}{2\pi}\,\Theta\prs{-\mu\mu'}\,\mathrm{e}^{-i\left(\mu\hmu+\mu'\hmu'\right)}\nonumber \\
 &  & \qquad\qquad\sum_{\mathbb{I}_{P}}\prod_{j\in\mathbb{I}_{P}}\cos\left(\frac{\hat{\mu}-\hat{\mu}'}{\sqrt{N}}B_{j}^{(m)}\right)\prod_{j\notin\mathbb{I}_{P}}\cos\left(\frac{\hat{\mu}+\hat{\mu}'}{\sqrt{N}}B_{j}^{(m)}\right).\label{eq:intermedio}\end{eqnarray}

The fraction of sets $\mathbb{I}_{P}$ with $n\leq m$ indexes $\ell\leq m$
is $\binom{m}{n}\binom{N-m}{P-n}/\binom{N}{P}$ and observing that
in the limit $P\leq N\to\infty$ with $P/N=p\leq1$ we have that \[
\lim_{P\leq N\uparrow\infty}\binom{N}{P}^{-1}\,\binom{N-m}{P-n}=p^{n}(1-p)^{m-n}.\]
 From equation (\ref{eq:intermedio}) we have that\begin{eqnarray*}
\mathfrak{d}^{(m)}(p) & = & \intoo\frac{\mathrm{d}\mu\,\mathrm{d}\hmu}{2\pi}\,\frac{\mathrm{d}\mu'\,\mathrm{d}\hmu'}{2\pi}\,\Theta\prs{-\mu\mu'}\,\mathrm{e}^{-i\left(\mu\hmu+\mu'\hmu'\right)}\\
 & \times & \sum_{n=0}^{m}\binom{m}{n}\, p^{n}(1-p)^{m-n}\,\cos\left(\frac{\hat{\mu}-\hat{\mu}'}{\sqrt{m}}\right)^{n}\cos\left(\frac{\hat{\mu}+\hat{\mu}'}{\sqrt{m}}\right)^{m-n}.\end{eqnarray*}
 By adding up the sum, opening up the cosines and applying the identity
$\Theta(ab)=\Theta(a)\Theta(b)+\Theta(-a)\Theta(-b),$ the expression
for the sensitivity gets reduced to\begin{eqnarray*}
\mathfrak{d}^{(m)}(p) & =2 & \sum_{n=0}^{m}\binom{m}{n}\,(1-2p)^{n}\left[\int\mathcal{D}(\mu,\hat{\mu})\,\cos(\hat{\mu}/\sqrt{m})^{m-n}\,\sin(\hat{\mu}/\sqrt{m})^{n}\right]^{2},\end{eqnarray*}
 where the notation $\int\mathcal{D}(\mu,\hat{\mu})\, f(\mu,\hat{\mu})$
stands for $(2\pi)^{-1}\int_{0}^{\infty}\mathrm{d}\mu\int_{-\infty}^{\infty}\mathrm{d}\hat{\mu}\,\mathrm{e}^{-i\mu\hat{\mu}}\, f(\mu,\hat{\mu}).$
The integrals to be solved are\begin{eqnarray*}
b_{0}^{m} & \equiv & \int\mathcal{D}\left(\eta,\hat{\eta}\right)\cos(\hat{\eta})^{m}\\
b_{n}^{m} & \equiv & \int\mathcal{D}\left(\eta,\hat{\eta}\right)\cos(\hat{\eta})^{m-2n}\sin(\hat{\eta})^{2n}\\
c_{n}^{m} & \equiv & \int\mathcal{D}\left(\eta,\hat{\eta}\right)\cos(\hat{\eta})^{m-(2n+1)}\sin(\hat{\eta})^{2n+1}.\end{eqnarray*}

Before computing the integrals observe that for all $A>0\;\mathrm{and}\; B\geq0$\begin{eqnarray}
\int\mathcal{D}\left(\eta,\hat{\eta}\right)\sin(A\hat{\eta}) & = & -\frac{i}{4\pi}\int_{0}^{\infty}\mathrm{d}\eta\int_{-\infty}^{\infty}\mathrm{d}\hat{\eta}\exp\left(-i\hat{\eta}\eta\right)\left[\exp\left(i\hat{\eta}A\right)-\exp\left(-i\hat{\eta}A\right)\right]\nonumber \\
 & = & -\frac{i}{4\pi}\int_{0}^{\infty}\mathrm{d}\eta\int_{-\infty}^{\infty}\mathrm{d}\hat{\eta}\left[\exp\left[-i\hat{\eta}(\eta-A)\right]-\exp\left[-i\hat{\eta}(\eta+A)\right]\right]\nonumber \\
 & = & -\frac{i}{2}\left[\Theta(A)-\Theta(-A)\right]\nonumber \\
 & = & -\frac{i}{2},\label{eq:integral-1}\end{eqnarray}
 similarly \begin{eqnarray}
\int\mathcal{D}\left(\eta,\hat{\eta}\right)\cos(A\hat{\eta})\cos(B\hat{\eta}) & = & \frac{1}{4}\left[\Theta(A+B)+\Theta(-A-B)+\Theta(A-B)+\Theta(-A+B)\right]\nonumber \\
 & = & \frac{1}{2}\label{eq:integral00}\end{eqnarray}
 and\begin{eqnarray}
\int\mathcal{D}\left(\eta,\hat{\eta}\right)\cos(A\hat{\eta})\sin(B\hat{\eta}) & = & -\frac{i}{4}\left[\Theta(A+B)-\Theta(A-B)+\Theta(-A+B)-\Theta(-A-B)\right]\nonumber \\
 & = & -\frac{i}{2}\Theta(B-A).\label{eq:integral01}\end{eqnarray}
 The first integral is (remember that \emph{m} is odd)\begin{eqnarray}
b_{0}^{m} & = & \int\mathcal{D}\left(\eta,\hat{\eta}\right)\cos(\hat{\eta})^{m}=\frac{1}{2^{m-1}}\sum_{k=0}^{(m-1)/2}\binom{m}{k}\int\mathcal{D}\left(\eta,\hat{\eta}\right)\cos[(m-2k)\hat{\eta}]\nonumber \\
 & = & \frac{1}{2^{m}}\sum_{k=0}^{(m-1)/2}\binom{m}{k}=\frac{1}{2}.\label{eq:b0}\end{eqnarray}

The second integral is \begin{eqnarray}
b_{n}^{m} & = & \int\mathcal{D}\left(\eta,\hat{\eta}\right)\cos(\hat{\eta})^{m-2n}\sin(\hat{\eta})^{2n}\nonumber \\
 & = & \int\mathcal{D}\left(\eta,\hat{\eta}\right)\frac{1}{2^{m-2n-1}}\sum_{k=0}^{(m-1)/2-n}\binom{m-2n}{k}\cos[(m-2(k+n))\hat{\eta}]\nonumber \\
 &  & \qquad\frac{1}{2^{2n}}\left\{ 2\sum_{j=0}^{n-1}(-1)^{n-j}\binom{2n}{j}\cos[(2n-2j)\hat{\eta}]+\binom{2n}{n}\right\} \nonumber \\
 & = & \frac{1}{2^{m-1}}\sum_{k=0}^{(m-1)/2-n}\binom{m-2n}{k}\,\sum_{j=0}^{n-1}(-1)^{n-j}\binom{2n}{j}+\frac{1}{2^{2n+1}}\binom{2n}{n}\nonumber \\
 & = & \frac{1}{2^{m-1}}\,\frac{2^{m-2n}}{2}\left[-\frac{1}{2}\binom{2n}{n}\right]+\frac{1}{2^{2n+1}}\binom{2n}{n}\nonumber \\
 & = & 0.\label{eq:bm}\end{eqnarray}

And the last integral is then\begin{eqnarray}
c_{n}^{m} & = & \int\mathcal{D}\left(\eta,\hat{\eta}\right)\cos(\hat{\eta})^{m-(2n+1)}\sin(\hat{\eta})^{2n+1}\nonumber \\
 & = & \int\mathcal{D}\left(\eta,\hat{\eta}\right)\cos(\hat{\eta})^{m-(2n+1)}\left[1-\cos^{2}(\hat{\eta})\right]^{n}\sin(\hat{\eta})\nonumber \\
 & = & \int\mathcal{D}\left(\eta,\hat{\eta}\right)\sum_{\ell=0}^{n}(-1)^{\ell}\,\binom{n}{\ell}\,\cos(\hat{\eta})^{m-(2n+1)+2\ell}\sin(\hat{\eta})\nonumber \\
 & = & \sum_{\ell=0}^{n}(-1)^{\ell}\,\binom{n}{\ell}\,\frac{(-i)}{2^{m-1+2(\ell-n)}}\left\{ \frac{1}{2}\,\binom{m-1+2(\ell-n)}{(m-1)/2+\ell-n}+\right.\nonumber \\
 &  & \qquad\left.+\sum_{k=0}^{(m-1)/2+\ell-n-1}\binom{m-1+2(\ell-n)}{k}\Theta\left[1-(m-1+2(\ell-n)-k)\right]\right\} \nonumber \\
 & = & -\frac{i}{2^{m-2n}}\sum_{\ell=0}^{n}\left(-\frac{1}{4}\right)^{\ell}\,\binom{n}{\ell}\,\binom{m-1+2(\ell-n))}{(m-1)/2+\ell-n}\nonumber \\
 & = & -\frac{i}{2^{m}}\,\binom{2n}{n}\,\binom{m-1-2n}{(m-1)/2-n}\,\binom{(m-1)/2}{n}^{-1}.\label{eq:am1}\end{eqnarray}

We have that, for all \emph{m} odd\begin{equation}
\mathfrak{d}^{(m)}(p)=\frac{1}{2}-\frac{1}{2}\sum_{n=0}^{(m-1)/2}a_{n}^{m}\,(1-2p)^{2n+1},\label{eq:dd}\end{equation}
 where\begin{equation}
a_{n}^{m}\equiv\frac{1}{4^{m-1}}{\displaystyle \,\binom{m}{2n+1}\,\left[\binom{2n}{n}\,\binom{m-1-2n}{(m-1)/2-n}\,\binom{(m-1)/2}{n}^{-1}\right]^{2}}.\label{eq:am}\end{equation}

Observe that $\mathfrak{d}^{(m)}(p)$ is concave in $p\in[0,\frac{1}{2}]$
(it is simply a sum of an affine plus concave functions) and $\mathfrak{d}^{(m)}(p)<\mathfrak{d}^{(m+2)}(p)$
for all $p\in(0,\frac{1}{2}).$ To demonstrate the latter we use that
$\mathfrak{d}^{(m)}(0)=0$ and $a_{s}^{m}>0\;\forall s$. Thus, from
(\ref{eq:dd}) at $p=0$ we have that\begin{equation}
\sum_{s=0}^{(m-1)/2}a_{s}^{m}=1\qquad\forall m\geq1.\label{eq:identity}\end{equation}
 Therefore\begin{eqnarray*}
a_{(m+1)/2}^{m+2} & = & \sum_{s=0}^{(m-1)/2}\left(a_{s}^{m}-a_{s}^{m+2}\right)\end{eqnarray*}
 simply by applying (\ref{eq:identity}) to $m$ and to $m+2$. Observe
that $(1-2p)^{n}<(1-2p)^{n'}$ for all $n>n'$ and $p\in(0,\frac{1}{2})$,
thus\begin{eqnarray*}
a_{(m+1)/2}^{m+2}\,(1-2p)^{m+2} & < & \sum_{s=0}^{(m-1)/2}\left(a_{s}^{m}-a_{s}^{m+2}\right)\,(1-2p)^{2s+1}\\
\sum_{s=0}^{(m+1)/2}a_{s}^{m+2}\,(1-2p)^{2s+1} & < & \sum_{s=0}^{(m-1)/2}a_{s}^{m}\,(1-2p)^{2s+1}\end{eqnarray*}
 and thus $\mathfrak{d}^{(m)}(p)<\mathfrak{d}^{(m+2)}(p)$.

In the large \emph{m} limit we have that \[
\lim_{m\uparrow\infty}\mathfrak{d}^{(m)}(p)=\frac{1}{\pi}\mathrm{arcos}(1-2p),\]
which is the expected result \citet{feller}.

\section{Optimal learning algorithm\label{sec:Optimal-learning}}

The basic ingredient to compute the optimal learning algorithm is
the joint probability distribution of the variables $\sigma_{\mathbf{B}}$,
\emph{h} and \emph{b.} Given that $\mathcal{P}(\sigma_{\mathbf{B}},h,b|m)=\Theta(\sigma_{\mathbf{B}}b)\,\mathcal{P}(h,b|m)$
we will start our inference task by computing the distribution of
the post-synaptic fields.\begin{eqnarray*}
\mathcal{P}(h,b|m) & = & \left\langle \delta\left(h-\mathbf{J}\cdot\mathbf{S}/|\mathbf{J}|\right)\,\delta\left(b-\mathbf{B}\cdot\mathbf{S}/|\mathbf{B}|\right)\right\rangle _{\mathbf{S}}\\
 & = & \int_{-\infty}^{\infty}\frac{\mathrm{d}\hat{h}}{2\pi}\,\mathrm{e}^{-i\hat{h}h}\int_{-\infty}^{\infty}\frac{\mathrm{d}\hat{b}}{2\pi}\,\mathrm{e}^{-i\hat{b}b}\left\langle \exp\left(i\hat{h}\frac{\mathbf{J}\cdot\mathbf{S}}{|\mathbf{J}|}+i\hat{b}\frac{\mathbf{B}\cdot\mathbf{S}}{|\mathbf{B}|}\right)\right\rangle _{\mathbf{S}}\end{eqnarray*}
 and assuming that $[\mathbf{B}]_{j}=\sqrt{\frac{N}{m}}\,\Theta(m+1-j)$
we can suppose that the student \emph{learns} this rule in such a
way that $[\mathbf{J}]_{j}\simeq\mathcal{J}\Theta(m+1-j)+\varepsilon_{j},$
where $\varepsilon_{j}\ll|\mathbf{J}|$ are i.i.d. variables. Therefore
\[
\frac{\mathcal{J}}{|\mathbf{J}|}=\frac{R}{\sqrt{m}}-\frac{\overline{\varepsilon}}{|\mathbf{J}|},\]
 where \emph{R} is the teacher-student overlap and $\overline{\varepsilon}\equiv\sum_{j=1}^{m}\varepsilon_{j}/m$.
Let us define the variables\[
\varphi_{j}\equiv\frac{\varepsilon_{j}-\overline{\varepsilon}}{|\mathbf{J}|},\]
 with the properties of $\sum_{j=1}^{m}\varphi_{j}=0$ and\[
\sum_{j>m}\frac{\varepsilon_{j}^{2}}{|\mathbf{J}|^{2}}=1-R^{2}-\sum_{j=1}^{m}\varphi_{j}^{2}.\]
 Thus the trace over the spin variables gives\begin{eqnarray*}
\left\langle \exp\left(i\hat{h}\frac{\mathbf{J}\cdot\mathbf{S}}{|\mathbf{J}|}+i\hat{b}\frac{\mathbf{B}\cdot\mathbf{S}}{|\mathbf{B}|}\right)\right\rangle _{\mathbf{S}} & = & \prod_{j=1}^{N}\frac{1}{2}\sum_{S=\pm1}\exp\left[i\left(\frac{\hat{h}[\mathbf{J}]_{j}}{|\mathbf{J}|}+\frac{\hat{b}[\mathbf{B}]_{j}}{|\mathbf{B}|}\right)S\right]\\
 & = & \prod_{j=1}^{m}\cos\left(\hat{h}\frac{\mathcal{J}+\varepsilon_{j}}{|\mathbf{J}|}+\frac{\hat{b}}{\sqrt{m}}\right)\,\prod_{j>m}\cos\left(\frac{\hat{h}\varepsilon_{j}}{|\mathbf{J}|}\right)\\
 & = & \prod_{j=1}^{m}\cos\left(\frac{\hat{h}R+\hat{b}}{\sqrt{m}}+\hat{h}\varphi_{j}\right)\,\prod_{j>m}\cos\left(\frac{\hat{h}\varepsilon_{j}}{|\mathbf{J}|}\right)\\
 & \simeq & \cos\left(\frac{\hat{h}R+\hat{b}}{\sqrt{m}}\right)^{m}\prod_{j=1}^{m}\left[1-\hat{h}\varphi_{j}\tan\left(\frac{\hat{h}R+\hat{b}}{\sqrt{m}}\right)+O\left(\varphi_{j}^{2}\right)\right]\\
 &  & \qquad\qquad\exp\left(-\frac{\hat{h}^{2}}{2}\sum_{j>m}\frac{\varepsilon_{j}^{2}}{|\mathbf{J}|^{2}}\right)\left[1+O\left(\sum_{j>m}\frac{\varepsilon_{j}^{4}}{|\mathbf{J}|^{4}}\right)\right]\\
 & \simeq & \cos\left(\frac{\hat{h}R+\hat{b}}{\sqrt{m}}\right)^{m}\exp\left(-\frac{1-R^{2}}{2}\,\hat{h}^{2}\right)+O\left(\sum_{j=1}^{m}\varphi_{j}^{2}\right).\end{eqnarray*}

Therefore, and using that \emph{m} is odd, \begin{eqnarray}
\mathcal{P}(h,b|m) & \simeq & \frac{1}{2^{m-1}}\sum_{k=0}^{(m-1)/2}\binom{m}{k}\int_{-\infty}^{\infty}\frac{\mathrm{d}\hat{h}}{2\pi}\,\frac{\mathrm{d}\hat{b}}{2\pi}\,\exp\left(-\frac{1-R^{2}}{2}\hat{h}^{2}-ih\hat{h}-ib\hat{b}\right)\cos\left[(m-2k)\frac{\hat{b}+R\hat{h}}{\sqrt{m}}\right]\nonumber \\
 & = & \mathcal{N}(h|Rb,1-R^{2})\,\frac{1}{2^{m}}\sum_{k=0}^{(m-1)/2}\binom{m}{(m-1)/2-k}\left[\delta\left(b-\mu_{k}\right)+\delta\left(b+\mu_{k}\right)\right],\label{eq:disthb}\end{eqnarray}
 where $\mu_{k}=(2k+1)/\sqrt{m}$ and $\mathcal{N}(x|\mu,\sigma^{2})$
is a Normal distribution in $x$, centered at $\mu,$ with variance
$\sigma^{2}.$

From (\ref{eq:disthb}) we can compute the joint distribution of the
variables \emph{h} and $\sigma_{\mathbf{B}}$\begin{equation}
\mathcal{P}(\sigma_{\mathbf{B}},h|m)=\frac{1}{2^{m}}\sum_{k=0}^{(m-1)/2}\binom{m}{(m-1)/2-k}\,\mathcal{N}(\sigma_{\mathbf{B}}h|R\mu_{k},1-R^{2}),\label{eq:probsigh}\end{equation}
 which implies that\begin{eqnarray}
\mathcal{P}(\phi|m) & = & \sum_{\sigma_{\mathbf{B}}=\pm1}\int_{-\infty}^{\infty}\mathrm{d}h\,\mathcal{P}(\sigma_{\mathbf{B}},h|m)\,\delta(\phi-\sigma_{\mathbf{B}}h)\nonumber \\
 & = & \frac{1}{2^{m-1}}\sum_{k=0}^{(m-1)/2}\binom{m}{(m-1)/2-k}\,\mathcal{N}(\phi|R\mu_{k},1-R^{2}).\label{eq:pfi}\end{eqnarray}
 The conditional probability of the field \emph{b} given $\sigma_{\mathbf{B}}$
and \emph{h} can be obtained from (\ref{eq:disthb}) and (\ref{eq:probsigh}).

It is a simple inference exercise to find the conditional distribution
of the field \emph{b} given the stability $\phi$\[
\mathcal{P}(b|\phi,m)=\frac{1}{2}\,\frac{\mathcal{N}(\phi|R|b|,1-R^{2})\sum_{k=0}^{(m-1)/2}\binom{m}{(m-1)/2-k}\delta(|b|-\mu_{k})}{\sum_{k=0}^{(m-1)/2}\binom{m}{(m-1)/2-k}\mathcal{N}(\phi|R\mu_{k},1-R^{2})}.\]

The conditional expectation of the field \emph{$|b|$} is\begin{eqnarray}
\left\langle |b|\right\rangle _{b|\phi,m} & = & \int_{-\infty}^{\infty}\mathrm{d}b\,|b|\,\mathcal{P}(b|\phi)\nonumber \\
 & = & \frac{\sum_{k=0}^{(m-1)/2}\binom{m}{(m-1)/2-k}\,\mu_{k}\,\mathcal{N}(\phi|R\mu_{k},1-R^{2})}{\sum_{k=0}^{(m-1)/2}\binom{m}{(m-1)/2-k}\,\mathcal{N}(\phi|R\mu_{k},1-R^{2})}.\label{eq:modb}\end{eqnarray}

\section{Derivation of (\ref{eq:im1m2mb})\label{sec:Derivation}}

In this Appendix we continue the development of (\ref{eq:int})\begin{eqnarray*}
\mathscr{I}_{m_{1},m_{2}} & = & \int\mathrm{d}\phi\,\mathcal{P}(\phi|m_{\mathbf{B}})\,\Upsilon(\phi|R,m_{1})\,\Upsilon(\phi|R,m_{2})\\
 & = & \frac{1}{2^{m_{\mathbf{B}}-1}}\sum_{k=0}^{(m_{\mathbf{B}}-1)/2}\binom{m_{\mathbf{B}}}{(m_{\mathbf{B}}-1)/2-k}\,\sum_{\ell_{1}=0}^{(m_{1}-1)/2}\,\sum_{\ell_{2}=0}^{(m_{2}-1)/2}\frac{R^{4}}{m_{1}m_{2}}\,\int_{-1}^{1}\mathrm{d}t_{1}\, t_{1}\,\int_{-1}^{1}\mathrm{d}t_{2}\, t_{2}\\
 &  & \qquad\int\mathrm{d}\phi\,\mathcal{N}(\phi|R\mu_{k})\,\mathcal{N}\left(\phi\left|\frac{R}{\sqrt{m_{1}}}(2\ell_{1}+1-t_{1})\right.\right)\,\mathcal{N}\left(\phi\left|\frac{R}{\sqrt{m_{2}}}(2\ell_{2}+1-t_{2})\right.\right),\end{eqnarray*}
 where all the Normal distributions have exactly the same variance
$1-R^{2}$. The integral over $\phi$ is simple and produces a bi-variate
Gaussian distribution in $t_{1}$ and $t_{2}$\begin{eqnarray}
\mathscr{I}_{m_{1},m_{2}} & = & \frac{1}{2^{m_{\mathbf{B}}-1}}\sum_{k=0}^{(m_{\mathbf{B}}-1)/2}\binom{m_{\mathbf{B}}}{(m_{\mathbf{B}}-1)/2-k}\,\sum_{\ell_{1}=0}^{(m_{1}-1)/2}\,\sum_{\ell_{2}=0}^{(m_{2}-1)/2}\frac{R^{2}}{\sqrt{m_{1}m_{2}}}\nonumber \\
 &  & \quad\int_{\mathscr{D}_{\mathbf{t}}}\mathrm{d}\mathbf{t}\, t_{1}\, t_{2}\,\mathcal{N}\left(\mathbf{t}\left|\mathbf{t}_{\ell_{1},\ell_{2},k};\mathbf{\Sigma}\right.\right)\label{eq:m1m2}\end{eqnarray}
 where $\mathbf{t}=(t_{1},t_{2})^{\mathsf{T}},$ $\mathscr{D}_{\mathbf{t}}\equiv(-1,1)\times(-1,1)$,
$\mathbf{t}_{\ell_{1},\ell_{2},k}=(\sqrt{m_{1}}\,\delta_{\ell_{1},k},\sqrt{m_{2}}\,\delta_{\ell_{2},k})^{\mathsf{T}}$
and \begin{eqnarray}
\delta_{\ell_{j},k} & \equiv & \frac{2\ell_{j}+1}{\sqrt{m_{j}}}-\frac{2k+1}{\sqrt{m_{\mathbf{B}}}},\label{eq:deltajk}\\
\mathbf{\Sigma} & \equiv & 2\,\frac{1-R^{2}}{R^{2}}\left(\begin{array}{cc}
m_{1} & \frac{1}{2}\sqrt{m_{1}m_{2}}\\
\frac{1}{2}\sqrt{m_{1}m_{2}} & m_{2}\end{array}\right).\label{eq:covariancia}\end{eqnarray}
 From (\ref{eq:approx1}) all the entries of the covariance matrix
(\ref{eq:covariancia}) are small, therefore all the distributions
are concentrated around $\mathbf{t}_{\ell_{1},\ell_{2},k}.$ Let $\mathbf{t}_{k}^{\star}$
be the vector that corresponds to the largest term in (\ref{eq:m1m2}).
Its components are

\begin{eqnarray}
\ell_{m_{j},k}^{\star} & \equiv & \left[\!\!\left|\sqrt{\frac{m_{j}}{m_{\mathbf{B}}}}\,\frac{2k+1}{2}-\frac{1}{2}\right|\!\!\right]\label{eq:lstar}\\
\sqrt{m_{j}}\,\delta_{m_{j},k}^{\star} & = & 2\ell_{m_{j},k}^{\star}+1-\sqrt{\frac{m_{j}}{m_{\mathbf{B}}}}\,(2k+1),\label{eq:delstar}\end{eqnarray}
 where $\left[\!\!\left|r\right|\!\!\right]$ is the closest integer
to $r\in\mathbb{R}$, thus $\sqrt{m_{j}}\,\delta_{m_{j},k}^{\star}\in(-1,1)$.
All the other vectors can be expressed as $\mathbf{t}_{\ell_{1},\ell_{2},k}=\mathbf{t}_{k}^{\star}+2\mathbf{n},$
where $\mathbf{n}=(n_{1},n_{2})^{\mathsf{T}},\, n_{j}=-\ell_{m_{j},k}^{\star},-\ell_{m_{j},k}^{\star}+1,\dots,-1,1,\dots,-\ell_{m_{j},k}^{\star}+(m_{j}-1)/2$.
We have that \begin{eqnarray*}
\mathscr{I}_{m_{1},m_{2}} & = & \frac{1}{2^{m_{\mathbf{B}}-1}}\sum_{k=0}^{(m_{\mathbf{B}}-1)/2}\binom{m_{\mathbf{B}}}{(m_{\mathbf{B}}-1)/2-k}\,\frac{R^{2}}{\sqrt{m_{1}m_{2}}}\\
 &  & \quad\left\{ \int_{\mathscr{D}_{\mathbf{t}}}\mathrm{d}\mathbf{t}\, t_{1}\, t_{2}\,\mathcal{N}\left(\mathbf{t}\left|\mathbf{t}_{k}^{\star};\mathbf{\Sigma}\right.\right)+\sum_{\mathbf{n}}\int_{\mathscr{D}_{\mathbf{t}}}\mathrm{d}\mathbf{t}\, t_{1}\, t_{2}\,\mathcal{N}\left(\mathbf{t}\left|\mathbf{t}_{k}^{\star}+2\mathbf{n};\mathbf{\Sigma}\right.\right)\right\} .\end{eqnarray*}
 Observe that the vectors $\mathbf{t}_{k}^{\star}$ are always strictly
inside the domain $\mathscr{D}_{\mathbf{t}}$. They can never be in
the boundary of the domain given that $m_{j}$ is odd then the argument
of the RHS of (\ref{eq:lstar}) is never in $\mathbb{Z}_{1/2}$ (which
would produce the largest possible value of $\sqrt{m_{j}}\,\delta_{m_{j},k}^{\star}$).
Thus, the largest contribution to the sum over $\mathbf{n}$ is of
$O\left[\exp\left(-\epsilon^{2}/\max\left\{ \lambda\in\mathrm{spec}(\mathbf{\Sigma})\right\} \right)\right],$
where $\epsilon\sim1-|\sqrt{m_{j}}\,\delta_{m_{j},k}^{\star}|>\frac{1}{2}$
and \[
\max\left\{ \lambda\in\mathrm{spec}(\mathbf{\Sigma})\right\} =\frac{1-R^{2}}{R^{2}}\left[m_{1}+m_{2}+\sqrt{m_{1}^{2}+m_{2}^{2}-m_{1}m_{2}}\right]\ll1,\]
according to (\ref{eq:approx1}). Within the same approximation error
we can suppose that the centre of the zero-th Normal distribution
is located inside the domain and sufficiently farther from the boundary.
Thus\begin{eqnarray*}
\mathscr{I}_{m_{1},m_{2}} & \simeq & \frac{1}{2^{m_{\mathbf{B}}-1}}\sum_{k=0}^{(m_{\mathbf{B}}-1)/2}\binom{m_{\mathbf{B}}}{(m_{\mathbf{B}}-1)/2-k}\,\frac{R^{2}}{\sqrt{m_{1}m_{2}}}\,\int_{\mathbb{R}^{2}}\mathrm{d}\mathbf{t}\, t_{1}\, t_{2}\,\mathcal{N}\left(\mathbf{t}\left|\mathbf{t}_{k}^{\star};\mathbf{\Sigma}\right.\right)+\\
 &  & \qquad+O\left(\exp\left[-\frac{\epsilon^{2}}{\max\left\{ \lambda\in\mathrm{spec}(\mathbf{\Sigma})\right\} }\right]\right).\end{eqnarray*}
 Thus\begin{eqnarray*}
\mathscr{I}_{m_{1},m_{2}} & \simeq & \frac{1}{2^{m_{\mathbf{B}}-1}}\sum_{k=0}^{(m_{\mathbf{B}}-1)/2}\binom{m_{\mathbf{B}}}{(m_{\mathbf{B}}-1)/2-k}\,\frac{R^{2}}{\sqrt{m_{1}m_{2}}}\\
 &  & \quad\int_{-\infty}^{\infty}\mathrm{d}t_{2}\, t_{2\,}\mathcal{N}\left(t_{2}\left|\delta_{m_{2},k}^{\star};2m_{2}\frac{1-R^{2}}{R^{2}}\right.\right)\\
 &  & \qquad\int_{-\infty}^{\infty}\mathrm{d}t_{1}\, t_{1\,}\mathcal{N}\left(t_{1}\left|\delta_{m_{1},k}^{\star}+\frac{1}{2}\sqrt{\frac{m_{1}}{m_{2}}}\,(t_{2}-\delta_{m_{2},k}^{\star});\frac{3}{2}m_{1}\frac{1-R^{2}}{R^{2}}\right.\right)\\
 & = & 1-R^{2}\left[1-\frac{1}{2^{m_{\mathbf{B}}-1}}\sum_{k=0}^{(m_{\mathbf{B}}-1)/2}\binom{m_{\mathbf{B}}}{(m_{\mathbf{B}}-1)/2-k}\,\delta_{m_{1},k}^{\star}\delta_{m_{2},k}^{\star}\right].\end{eqnarray*}


\begin{thebibliography}{10}
\bibitem{rumelhart}D. E. Rumelhart and J. L. MacClelland, \emph{Parallel
distributed processing }vol I, MIT Press, Cambridge, MA, 1986.

\bibitem{papert}M. Minsky and S. A. Papert, \emph{Perceptrons: An
Introduction to Computational Geometry}, MIT Press, Cambridge, MA,
expanded edition, 1988/1989.

\bibitem{engel}A. Engel and C. Van den Broeck, \emph{Statistical
Mechanics of Learning,} Cambridge Univ. Press, 2001.

\bibitem{church}A. Church, Am. J. Math. \textbf{58}, 345 (1936).

\bibitem{turing0}A. M. Turing, Proc. London Math. Soc. \textbf{42},
230 (1937).

\bibitem{hartmanis}J. Hartmanis and R. E. Stearns, \textit{\emph{Trans.
Am. Math. Soc.}} \textbf{117}, 285 (1965).

\bibitem{kolmogorov}A. N. Kolmogorov, \emph{P}roblems of Information
and Transmission \textbf{1}, 1 (1965).

\bibitem{ming}M. Li and P. Vitányi, \emph{An introduction to Kolmogorov
Complexity and its Applications}, 2nd edition, Springer, 1997.

\bibitem{fra-0}L. Franco and S. Cannas, Neural Computation \textbf{12},
2405 (2000).

\bibitem{fra-1}L. Franco and S. Cannas, \textit{\emph{Physica A}}
\textbf{332}, 337 (2004).

\bibitem{fra-2}L. Franco and M. Anthony, Proc. of the IEEE Int. Joint
Conf. on Neural Networks (Budapest, Hungary) p~973, 2004.

\bibitem{ton}A. C. C. Coolen, R. K\"uhn and P. Sollich, \emph{Theory
of Neural Information Processing Systems}, Oxford Univ. Press, 2005.

\bibitem{canning}A. Canning and E. Gardner, J. Phys. A \textbf{21},3275
(1988).

\bibitem{bouten}M. Bouten, A. Engel, A. Komoda and R. Serneels, J.
Phys. A \textbf{23}, 4643 (1990).

\bibitem{jort}D. Boll\'e and J. van Mourik, J. Phys. A \textbf{27},
1151 (1994).

\bibitem{khulman}P. Kuhlmann and K. R. M\"uller, J. Phys. A \textbf{27},
3759 (1994).

\bibitem{lopez}B. L\'opez and W. Kinzel, J. Phys. A \textbf{30},
7753 (1997).

\bibitem{malzahn}D. Malzahn, Phys. Rev. E \textbf{61}, 6261 (2000).

\bibitem{kalai}G. Kalai, Adv. in Appl. Math. \textbf{29}, 412 (2002).

\bibitem{odonnell}R. W. O'Donnell, \emph{Computational Applications
of Noise Sensitivity}, MIT Thesis, 2003.

\bibitem{bs}J. Stoer and R. Bulirsch, \emph{Introduction to numerical
analysis}, Springer Verlag, New York, 1980.

\bibitem{caki}O. Kinouchi and N. Caticha, Phys. Rev. E \textbf{54},
R54 (1996).

\bibitem{feller}W. Feller, \emph{An Introduction to Probability Theory
and Its Applications}, Second Edition, John Wiley, New York, 1957.

\end{thebibliography}
\end{document}